\documentclass[twocolumn,aps,prb,english,epsfig,groupedaddress]{revtex4}
\usepackage{graphicx}
\usepackage{amsmath}
\usepackage[latin1]{inputenc}
\usepackage{latexsym}
\usepackage{ulem}
\usepackage{color}

\begin{document}

\title{Dynamical properties of a 3-dimensional diluted Heisenberg model.}
\author{Akash Chakraborty$^{1}$ and Georges Bouzerar$^{1,2}$}
\affiliation{
$1.$ Institut N\'eel, CNRS, d\'epartement MCBT, 25 avenue des Martyrs, B.P. 166, 38042 Grenoble Cedex 09, France \\
$2.$ Jacobs University Bremen School of Engineering and Science, Campus Ring 1, D-28759 Bremen, Germany\\
}

\date{\today}
\begin{abstract}
We study the magnetic excitation spectrum in 3D diluted ferromagnetic nearest neigbour systems down to the percolation threshold.
The disorder effects resulting from the dilution are handled accurately within Self-Consistent Local RPA approach.
The calculations are performed using relatively large systems containing typically 20000 localised spins, a systematic average over many configurations of disorder is performed. We analyze in details the change in the magnon spectrum and magnon density of states as we increase the dilution. The zone of stability of the well defined magnon modes is shown to shrink drastically as we approach the percolation threshold.
 We also calculate the spin stiffness which appear to vanish at the percolation threshold exactly.
 A comparison with available data, based on a different theoretical approach, is also provided. We hope that this study will motivate new experimental studies based on inelastic Neutron Scattering measurements.
\end{abstract}

\maketitle

Over the last few decades interest in disordered magnetic systems, like
transition metal alloys \cite{abrikosov,moruzzi,james,turek}, diluted magnetic semiconductors as Mn doped III-V compounds \cite
{matsukura,ohno2,edmonds,blinowski,akai,koshihara,wang,sato,wierzbowska},
$d^{0}$ materials\cite{stoneham,elfimov,osorio,coey1,young,bouzerar,maca} and manganites \cite{hennion,ye,coey2,karmakar,motome} has considerably
increased. Most of these materials have attracted a great deal of attention from both the experimental as well as the theoretical point of view due to their potential in spintronics applications. Inelastic neutron scattering is the most direct experimental method to study the magnetic excitations in these materials. However, theoretical studies require tools which can treat both disorder and thermal fluctuations in a reliable manner.
Different theoretical methods have been performed/developed to calculate magnetic properties and the transition temperatures. Among the appropriate tools one can first quote the classical Monte Carlo method which is essentially exact. However to our knowledge no studies on the magnetic excitation spectrum
in dilute 3d Heisenberg model has been performed, probably because of the cost both in memory and
CPU time. Most of the existing studies within classical Monte Carlo
method deal with the equilibrium properties of the Ising model \cite{pawley,landau,reis}and there are a few focusing on the 3D diluted nearest neighbor Heisenberg model\cite{gordillo}. There is an alternative approach based on finite temperature Green's functions known as Self-Consistent Local Random Phase approximation (SC-LRPA). It was shown to be reliable to handle dilution/disorder effects and thermal fluctuations \cite{bouzerar0,bouzerar1,bouzerar2}. In the SC-LRPA theory disorder/dilution is treated exactly while thermal fluctuations are treated within random phase approximation. This method has several advantages, it is semi-analytical, fast, and allows to reach very large systems. The aim of the present manuscript is to provide a detailed study of the magnetic excitation spectrum in the dilute ferromagnetic nearest-neighbor Heisenberg Hamiltonian.

Let us now present a short summary of the SC-LRPA formalism. The Hamiltonian describing N$_{imp}$ interacting quantum/classical spins randomly distributed on a lattice of N sites is given by the diluted random Heisenberg model,
\begin{eqnarray}
H=-\sum_{i,j} J_{ij}p_{i}p_{j} {\bf S}_{i}\cdot {\bf S}_{j}
\label{Hamiltonian}
\end{eqnarray}
where the couplings J$_{ij}$ are assumed to be given and are restricted to nearest neighbor only (J). The sum {ij} runs over all sites and the random variable p$_i$ is 1 if the site is occupied by a spin otherwise it is 0. In the present work, the calculations are performed at $T=0$ K, the
 quantum/classical nature of the spin is irrelevant. Thus we assume classical spins with
$\mid{\bf S}_{i}\mid$=1.

We introduce the retarded Green's functions G$_{ij}^c(\omega)$=$\int_{-\infty}^{+\infty}$
 G$_{ij}^c(t)e^{i{\omega}t}dt$ where G$_{ij}^c(t)$= -i$\theta(t)\langle [S_i^+(t),S_j^-(0)]\rangle
$, which describe the transverse spin fluctuations and the subscript `$c$' corresponds to the
configuration of disorder. After performing the Tyablicov decoupling of the higher
order Green's functions which appear in the equation of motion of G$_{ij}^c(\omega)$ \cite{bouzerar2}, we obtain
\begin{eqnarray}
({\omega}{\bf I} -{\bf H}_{eff}^c){\bf G}^c={\bf D}
\label{matrix}
\end{eqnarray}
where ${\bf H}_{eff}^c$,${\bf G}^c$ and ${\bf D}$ are $ N_{imp}\times N_{imp}$ matrices.

The effective Hamiltonian matrix elements are (${\bf H}$$_{eff}^c$)$_{ij}$ = -$\langle S_i^z
\rangle$ J$_{ij}+ \delta_{ij}\sum_l \langle S_l^z\rangle$J$_{lj}$ and D$_{ij}$=2$\langle S_i^z
\rangle \delta_{ij}$. For a given temperature and disorder configuration, the local
magnetizations $\langle {S_i^z} \rangle$ ($i=1,2,....N_{imp}$) have to be calculated self-consistently at each temperature. The local magnetization is evaluated from the Callen expression \cite{callen}, which relates the local Green's function at site i to the local magnetization at this site,
\begin{eqnarray}
\langle S_i^z \rangle=\frac{(S-\Phi_i)(1+\Phi_i)^{2S+1}+(1+S+\Phi_i){\Phi_i}^{2S+1}}{(1+\Phi_i)^{2S+1}-\Phi_i^{2S+1}}
\label{Callen}
\end{eqnarray}
The local effective magnon occupation number is given by $\Phi_i=-\frac{1}{2\pi\langle{S_i^z}\rangle}\int_{-\infty}^{+\infty}\frac{\Im G_{ii}(\omega)}
{exp(\omega/kT)-1}d\omega$.

Despite the matrix being non-Hermitian, the spectrum is real and positive at each temperature in
the ferromagnetic phase. A negative eigenvalue would indicate an instability of the ferromagnetic
phase. H$_{eff}^c$ is non-Hermitian (real non-symmetric) but it has the property to be
biorthogonal\cite{biorth}. Hence one has to define the right and left eigenvectors of H$_{eff}^c$ denoted by $|\Psi_\alpha^{R,c}\rangle$ and  $|\Psi_\alpha^{L,c}\rangle$ respectively, both associated to the same eigenvalue $\omega_\alpha^c$. After inserting the L and R eigenvectors in eq.(2), the retarded Green's functions can be written as
\begin{eqnarray}
G_{ij}^c(\omega)=\sum_\alpha \frac{2\langle{S_j^z}\rangle}{\omega-\omega_\alpha^c+i\epsilon}
\langle i|\Psi_\alpha^{R,c}\rangle \langle\Psi_\alpha^{L,c}|j\rangle
\label{GF}
\end{eqnarray}
The averaged Fourier transform is given by $\bar{G}({\bf q},\omega)=\langle \frac{1}{N_{imp}}\sum_{ij}
e^{i{\bf q}(r_i-r_j)}G_{ij}^c(\omega)\rangle_c $, where the notation $\langle.....\rangle_c$
denotes the average over the disorder configurations. The dynamical spectral function is given by
$\bar{A}({\bf q},\omega)=-\frac {1}{\pi\langle\langle{S^z}\rangle\rangle}$Im$\bar{G}({\bf q},\omega)
$, where $\langle\langle{S^z}\rangle\rangle=\frac {1}{N_{imp}}\sum_i\langle{S_i^z}\rangle$ is the
total magnetization averaged over all spin sites. The spectral function $\bar{A}({\bf q},
\omega)$ provides direct access to the magnetic excitation spectrum, it is also directly accessible by
Inelastic Neutron Scattering experiments.

One can also calculate the magnon density of states which is given by $\rho(\omega)=\frac {1}{N_{imp}}\sum_i\rho_i(\omega)$, where $\rho_i(\omega)=-\frac{1}{2\pi\langle S_i^z \rangle}$Im$G_{ii}(\omega)$, is the local magnon DOS. Here we have calculated the magnetic excitation spectrum for a simple cubic lattice as a function
of the concentration of the spins $x=\frac{N_{imp}}{N}$. The calculations are
performed at T=0 $K$, where the matrix H$_{eff}$ is real symmetric and hence L and R eigenvectors
are identical. Since the couplings used in this case are all ferromagnetic, there is no frustration
and the system exhibits long range ferromagnetic order beyond the percolation threshold. \begin{figure}
\centering
\includegraphics[width=6cm,angle=-90]{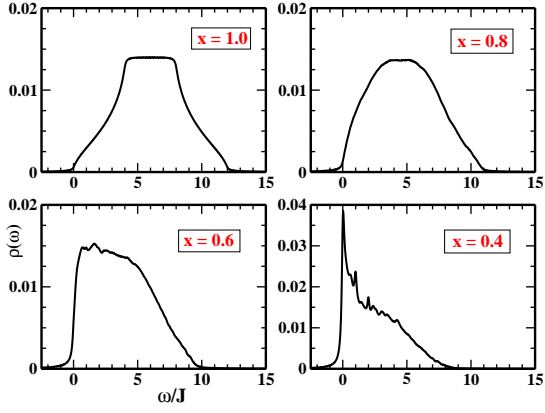}
\caption{(Color online) Magnon density of states $\rho(\omega)$ as a function of the energy $\omega$ for different concentration of localised spins $x$.The energy axis (x axis) is in $\omega/J$.}
\label{Fig1}
\end{figure}

In Fig.\ref{Fig1} we have plotted the magnon density of states (DOS) as a function of the energy, for different concentrations of localised spins ($x$=1.0 corresponds to the clean case). As the dilution increases, we observe (i) a significant change in the shape of the magnon DOS, (ii) a reduction of the  bandwidth and (iii) an increase in the low energy DOS. As we approach the percolation threshold which is located at $x_c$=0.31\cite{kirk}, we observe for the lowest concentration a sharp peak developing at $\omega$=0. This feature is attributed to the formation of several isolated clusters which have their own zero energy modes. Additionally we observe at higher energy, peak structures located at $\omega$=J and 2J respectively. These peaks are due to the non-zero eigenmodes of these isolated clusters.
Note that the DOS does not provide any information on the nature (extended/localized) of
the magnon modes. This would require a more careful analysis and study, either using the calculation of the inverse participation ratio (IPR) or the typical magnon DOS\cite{nikolic} which provide a direct access to the ``mobility edge'' separating the localized modes from the extended ones. The typical magnon DOS is given by $\rho_{typ}(\omega)$ = exp (${\langle}$ln ${\rho_i(\omega)\rangle}_c$) , where $\langle.....\rangle_c$ represents the average over disorder configurations and ${\rho_i(\omega)}$ is the local magnon DOS. We expect the mobility edge to become zero at the percolation threshold exactly, for a review see  \cite{review_fracton}.

\begin{figure}[htbp]
\centering
\includegraphics[width=7.0cm,angle=-90]{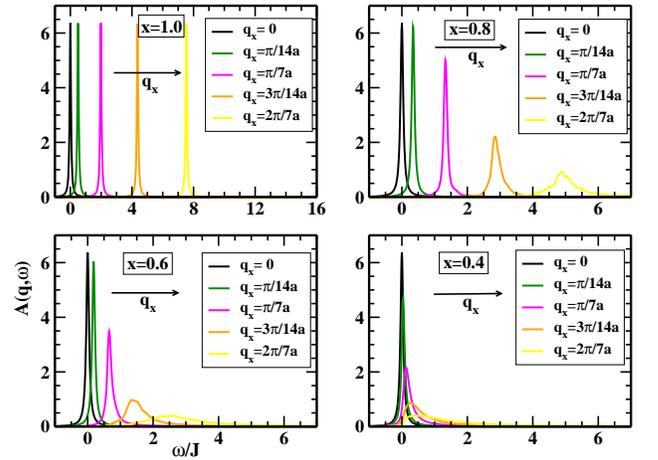}
\caption{(Color online) Spectral function $A({\bf q},\omega)$ as a function of the energy $\omega$ in the (1,0,0) direction for different values of $q_x$, with varying concentration of localised spins ($x$).The energy axis (x axis) is in $\omega/J$}
\label{Fig2}
\end{figure}

 Fig.\ref{Fig2} shows the spectral function $\bar{A}({\bf q},\omega)$ as a function of energy for different values
of the momentum ${\bf q}$ in the (100) direction. We have shown the spectral function for different concentrations of spins $x$, with $x$=1.0 corresponding to the clean case. The localised spins are randomly
distributed over a simple cubic lattice of size L$^3$ (L=28) and we have used periodic boundary
conditions. For x=0.8, the system contains approximately 17000 spins. We have examined the
finite-size effect by comparing with the results for L=20 and L=24, and concluded that L=28 is
large enough for the following discussions. The average over
disorder was done for 50 configurations. We have also checked for the necessary number of random
configurations and found that 20 configurations are sufficient for L=28 system.  The number of disorder configurations should be
sufficient in order to extract the correct spin stiffness. We can clearly see from the figure
that well defined excitations exist only for small values of ${\bf q}$. As the momentum increases
the peak becomes broader and develops a tail extending towards higher energies. As we get closer
to the percolation threshold, the magnetic excitation peaks become strongly asymmetric. Similar
asymmetric peaks were also observed in the magnetic excitation spectrum for the diluted
antiferromagnet Mn$_x$Zn$_{1-x}$F$_2$ \cite{uemura}. This increase in asymmetry with the momentum  corresponds to a crossover from propagating spin waves to localized excitations (fractons)\cite{orbach1,orbach2}.

\begin{figure}[htbp]
\centering
\includegraphics[width=8cm,angle=0]{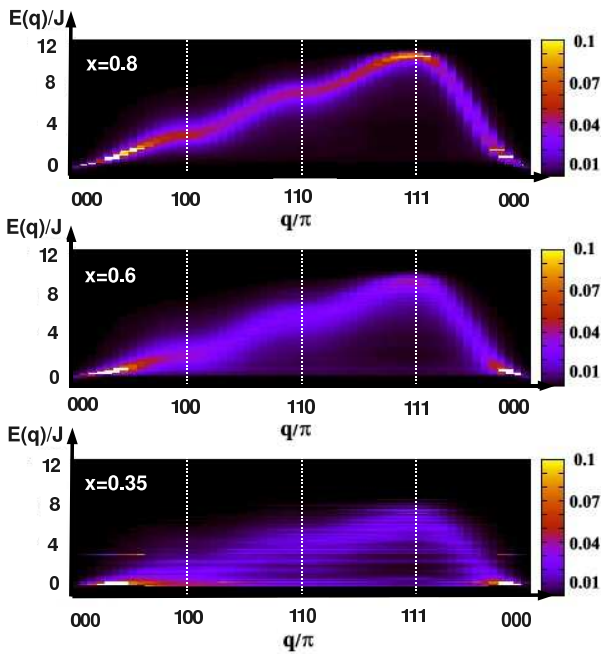}
\caption{(Color online) Spectral function $\bar{A}({\bf q},\omega)$ in the $(q,\omega)$ plane for different concentration of localised spins ($x$). (The size of the simple cubic lattice is
L=32).}
\label{Fig3}
\end{figure}

\begin{figure}[htbp]
\centering
\includegraphics[width=7cm,angle=-90]{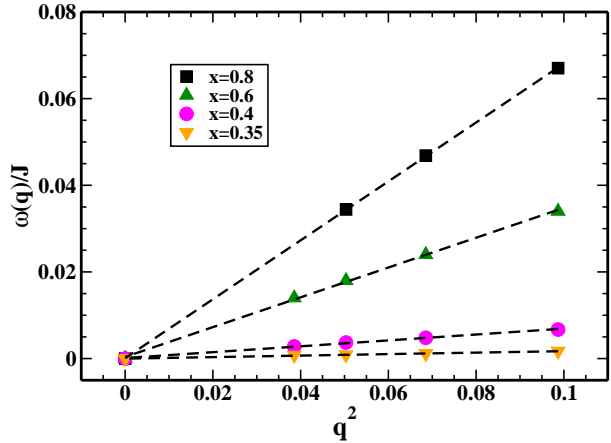}
\caption{(Color online) Magnon energy $\omega(q)$ (in $\omega/J$) as a function of $q²$ for different concentration of localised spins ($x$). The size of the simple cubic lattice varies from L=20,24,28 and 32.}
\label{Fig4}
\end{figure}

In the next figure, Fig.\ref{Fig3}, we have plotted the spectral function in the $(q,\omega)$ plane, corresponding to three different concentrations of spins. The spectral function is shown for the entire Brillouin zone. Well defined excitations can be observed
in the dilute case only in a restricted region of the Brillouin zone close to the $\Gamma$
point ($q$=(000)). Similar results were observed for the excitation spectrum of the diluted
magnetic semiconductor Ga$_{1-x}$Mn$_{x}$As \cite{bouzerar3}. From the figures we conclude that well defined excitations cease to exist as we go to higher dilutions and the spectrum shows broadening close to the
percolation threshold due to the formation of localized modes.

\begin{figure}[htbp]
\centering
\includegraphics[width=7cm,angle=-90]{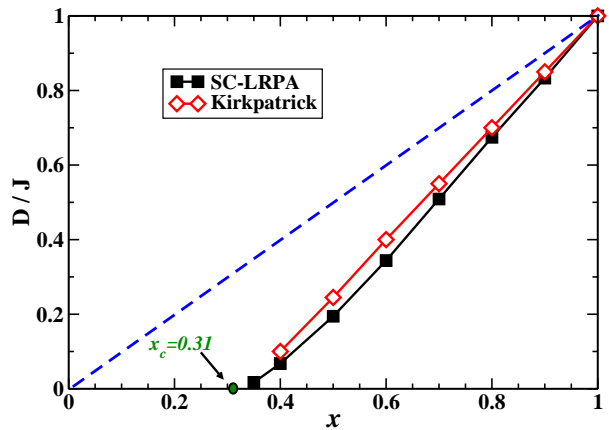}
\caption{(Color online) Spin stiffness D (in D/J) as a function of the concentration of the localised spins ($x$).
The diamonds represent the data from ref.\cite{kirk}.The squares are calculated within SC-LRPA.
The dashed line is the Virtual Crystal Approximation value. ($x_c$ is the percolation threshold).}
\label{Fig5}
\end{figure}

The spin wave energy $\omega(q)$ for different concentrations of spins as a function of $q²$ is shown in Fig.\ref{Fig4}. Actually the energy of the first peak in $A({\bf q},\omega)$ in the (100) direction is shown as a function of $q²$. The system size varies from L=20 to L=32. As can be seen from the figure for small momentum, $\omega(q)\approx D(x)q^2$. The slope of these curves gives the spin stiffness $D(x)$ for different $x$. We observe a strong decrease of the slope as we approach the percolation threshold. Note that, the perfect linear behaviour
of $\omega(q)$ as a function of q$^2$ vindicates the fact that the average over disorder configurations was obviously sufficient. In Fig.\ref{Fig5} we have plotted the spin stiffness $ D(x)$ as a function of the concentration of the localised spins $x$. We find that the spin stiffness is almost linear from $x=1.0$ down to $x=0.5$ and only forms a concave toe close to the percolation. We have also shown the spin stiffness obtained by Kirkpatrick \cite{kirk}.The author has used site percolation statistics on random resistor networks, as a natural generalization of lattice models, and obtained a relation between the site percolation probability $P^{(s)}(x)$, the conductance $G(x)$ and the spin stiffness coefficient $D(x)$. We find that our results are in very good agreement with those of ref.\cite{kirk}. Note that, the method we have used here is entirely different since the spin-stiffness is extracted directly from the curvature of the magnons excitations. We have also plotted the Virtual Crystal Approximation (VCA) value which is given by $D^{VCA}(x)= x D(x=1)$. As we decrease the concentration of localized spins the difference between the VCA  and the SC-LRPA value increases significantly. Below $x=0.5$ the  $D^{VCA}(x)$ drastically overestimates the calculated values. Note that our spin stiffness curve is very different from that obtained in the case of optimally annealed samples of the diluted magnetic semiconductor Ga$_{1-x}$Mn$_{x}$As \cite{bouzerar3}. This difference of behavior is due to the fact that in the case of  Ga$_{1-x}$Mn$_{x}$As the concentrations were much smaller and that the couplings were rather extended and spin concentration dependant.

In conclusion, we have calculated the magnetic excitation spectrum for a dilute and disordered
ferromagnetic system by using an approach based on the SC-LRPA, which allows us to treat disorder and
thermal fluctuations in a reliable manner. We see that a proper treatment of the disorder leads to unusual excitation spectrum compared to what is usually observed in non dilute systems. We have also calculated the spin stiffness, without adjusting parameters, as  a function of the spin concentration and observed  that our results are in very good agreement with previous studies for a similar model. Earlier the SC-LRPA was proved to be reliable for the determination of the Curie temperature \cite{bouzerar1}. Here we have shown its accuracy and efficiency for the calculation of the magnetic excitation spectrum. We hope that this study will stimulate new experimental measurements like the dynamical spectral function and spin stiffness as a function of the localised spin concentration.

\acknowledgments
We would like to thank Richard Bouzerar and Arnaud Ralko  for interesting discussions and remarks. AC would also sincerely thank the RTRA Nanosciences Fondation for extending their valuable support to this project. AC would also thank Soumen Mandal for helping with some technical details regarding the manuscript.

\end{document}